# Necklace-like pattern of vortex bound states

Zhiyong Hou[1†], Kailun Chen[1†], Wenshan Hong[2,3†], Da Wang[1†], Wen Duan[1], Huan Yang[1*], Shiliang Li[2,4,5], Huiqian Luo[2,5], Qiang-Hua Wang[1*], Tao Xiang[2,4,6], Hai-Hu Wen[1*]

[1]National Laboratory of Solid State Microstructures and Department of Physics, Collaborative Innovation Center of Advanced Microstructures, Nanjing University, Nanjing 210093, China

[2]Beijing National Laboratory for Condensed Matter Physics, Institute of Physics, Chinese Academy of Sciences, Beijing 100190, China

[3]International Center for Quantum Materials, School of Physics, Peking University, Beijing 100871, China

[4]School of Physical Sciences, University of Chinese Academy of Sciences, Beijing 100190 China

[5]Songshan Lake Materials Laboratory, Dongguan, Guangdong 523808, China

[6]Beijing Academy of Quantum Information Sciences; Beijing, China

*Corresponding author. Email: huanyang@nju.edu.cn (H.Y.); qhwang@nju.edu.cn (Q.-H.W.); hhwen@nju.edu.cn (H.-H.W.)

†These authors contributed equally to this work.

**Abstract:** Vortex is a topological defect in the superconducting condensate when a magnetic field is applied to a type-II superconductor, as elucidated by the Ginzburg-Landau theory. Due to the confinement of the quasiparticles by a vortex, it exhibits a circular shaped pattern of bound states with discrete energy levels, as predicted by the Caroli-de Gennes-Matricon theory in 1964. Here, however, we report a completely new type of vortex pattern which is necklace-like in an iron-based superconductor $KCa_2Fe_4As_4F_2$. Our theoretical analysis shows that this necklace-like vortex pattern arises from selective off-shell interference between vortex bound states of opposite angular momenta in the presence of rotational symmetry breaking due to disorders. This fascinating effect can be observed in a system with a small Fermi energy and wave vector, conditions fortuitously met in our samples. Our results not only disclose a novel vortex structure but also provide insights into comprehending the physics of the superconducting condensate.

**One-Sentence Summary:** A new type of vortex pattern with a necklace-like shape has been observed in the iron-based superconductor $KCa_2Fe_4As_4F_2$.

**Main Text:** In a type-II superconductor, vortices with quantized magnetic flux $\Phi_0 = h/2e$ (with $h$ the Planck constant and $e$ the elementary electric charge) are formed when the magnetic field penetrates the superconductor. Bogoliubov quasiparticles (QPs) exist in the circularly symmetric vortex core with a radius of about the coherence length $\xi \approx \hbar v_F/\pi\Delta_0$ with $v_F$ the Fermi velocity and $\Delta_0$ the bulk superconducting gap value. Based on the Ginzburg-Landau (GL) theory (*1*), the superconducting gap $\Delta$ is zero at the vortex center. It gradually increases along the radial direction and finally reaches $\Delta_0$. Since the surrounding superconducting region is protected by the energy gap, the vortex core can be considered as a circular quantum well for QPs in a clean superconductor. As a result, vortex bound states (VBS) or the so-called Caroli-de Gennes-Matricon (CdGM) states will be formed (*2*). Based on a simplified analytic solution to the Bogoliubov-de Gennes (BdG) equations by assuming a linear radial ($r$) variation of $\Delta$, it is predicted that the discrete energy levels of the VBS should appear at $E_\mu \approx \mu\Delta_0^2/E_F$ (the coefficient $\mu = \pm1/2, \pm3/2, \pm5/2, \ldots$) with $E_F$ the Fermi energy (*2-4*). In most superconductors, $\Delta_0$ is very small compared to $E_F$, and therefore, the energy interval $\Delta_0^2/E_F$ between neighboring levels of VBS is too small to distinguish. However, in some iron-based superconductors, $E_F$ is comparable with $\Delta_0$, making them ideal platforms for detecting the discrete VBS. The clear observation of discrete VBS, especially the high-order ones, was achieved in FeTe$_{0.55}$Se$_{0.45}$ (*5,6*) and FeSe monolayer (*7*) by scanning tunneling microscopy (STM). In these studies, the ratio of the lowest bound state energies is close to the original prediction of 1:3:5. However, this ratio is found to deviate from the widely believed value in KCa$_2$Fe$_4$As$_4$F$_2$ (K12442, $T_c \approx 33.5$ K) (*8*); theoretical analysis indicates that the deviation originates from the nonlinear relationship between $\Delta$ and $r$ in the extreme quantum limit condition (*9-12*). In addition, spatial oscillations of the bound states can be observed in the radial direction of the vortex core (*8-10,13,14*). Meanwhile, in some iron-based superconductors with topological nontrivial band structures, Majorana zero mode, as a particular VBS with $\mu = 0$, can be observed in the vortex cores (*15-18*), which may be used for topological quantum computing. Usually, the vortex core image and VBS keep the rotational symmetry showing a round ring shape of continuous intensity of density of states (DOS).

Here, we report a new type of VBS in K12442 with clear oscillations along the angular direction of the ring featuring the DOS, forming a necklace-like vortex pattern. This has never been predicted by any previous theoretical studies. Based on theoretical calculations involving suitable disorder scattering, we found that this novel phenomenon can be interpreted by the quasiparticle interference (QPI) among different CdGM states with the dominant contributions from those with opposite angular momenta. Conversely, such an interference phenomenon provides a new approach to detect the phase information of the CdGM states, which has long been overlooked in the research of VBS.

# Results

## Observation of the necklace-like vortex pattern

For a superconductor with an isotropic gap, the discrete energy levels of VBS are schematically shown in Fig. 1A based on the calculated results (fig. S1). The calculation is based on the exact diagonalization of the BdG Hamiltonians (see supplementary materials) in the extreme quantum limit, and the discrete VBS can be seen at different energies. Figure 1B shows the intensity of the DOS at zero bias, which is induced by the thermal broadening of the lowest levels of VBS. When the energy is not zero, VBS manifest themselves as continuous rings of the DOS along the angular direction (Fig. 1C). As the energy increases, higher levels of VBS become dominant and the DOS ring expands its radius (Fig. 1D). In addition, there is a spatial oscillation of the DOS along the radial direction with the period of $\pi/k_F$ (*9,13,14*), where $k_F$ is the Fermi wave vector, and one can see the secondary rings outside the main DOS rings (Fig. 1, C and D).

Although K12442 is a multi-band superconductor (*19*), the major contribution to the surface DOS is given by the smallest hole-like α pocket near Γ point with $k_F \approx 0.1\pi/a_0$ from our previous work (*20*). Discrete VBS can be observed in the vortex core (figs. S3 and S4) because the extreme quantum limit is satisfied (*8*). Vortices can be imaged by differential conductance (d$I$/d$V$) mapping at different energies under a magnetic field of 2 T (Fig. 1, E to G, fig. S2). The lowest VBS ($\mu = \pm 1/2$) peak can be observed at about ±1.0 meV on the spectrum measured at the vortex center (fig. S2), while the higher-order VBS peaks can be observed when moving away from the core center. A full gap feature with $\Delta_0 = 5.2$ meV can be clearly seen when the location is far away from the vortex center, indicating nodeless superconductivity of K12442 (*20,21*). Comparing the experimental (Fig. 1, F and G) and the calculated (Fig. 1, C and D) images of a vortex core at different energies, one can see that the radial variations of DOS show similar behaviors. However, the distribution of DOS along the angular direction appears to strongly deviate from the theoretical result: the calculated DOS exhibits circular rings with uniform and continuous DOS at finite energies below $\Delta_0$, while the measured DOS shows a periodic oscillation in angular direction and the vortex core shows a necklace-like pattern. The oscillation of DOS arising from the VBS behaves as alternative bright and dark spots along the necklace-like pattern (Fig. 1, F and G, fig. S2). The average period of these oscillations is about 3.8 nm, which is close to the value of $\pi/k_F$ in K12442 (*20*).

The necklace-like pattern has also been observed in other vortex cores (figs. S3 and S4). To further study this novel VBS, we measured the d$I$/d$V$ mapping in a large area (Fig. 2A and fig. S5) to see whether this phenomenon is universal in the system. Under a magnetic field of 2 T, vortices are clearly observed in this area (Fig. 2, B to D). However, they do not form a perfect hexagonal lattice, which might be attributed to the vortex pinning effect. In addition, some vortices seem to show a little squarish shape, especially at high energies, which may be due to a slightly fourfold anisotropy of the Fermi velocity and/or the superconducting gap. But this seems not to influence the oscillation behavior of DOS on the ring, nor the average distance between the neighboring spots (fig. S5). It is evident that the necklace-like pattern exists for each vortex, and does not depend on the slightly different shape of the vortex.

**Interference of the VBS along the angular direction**

The angular oscillation behaviors of the VBS motivate us to consider the interference effect between different CdGM states. For convenience, we use integer quantum number $l$ ($l = \mu - 1/2$) to label the angular momentum of the standard CdGM states

$$\begin{bmatrix} u_l(\boldsymbol{r}) \\ v_{l+1}(\boldsymbol{r}) \end{bmatrix} = \begin{bmatrix} \psi_l(r)e^{-il\varphi} \\ \psi_{l+1}(r)e^{-i(l+1)\varphi} \end{bmatrix} \quad (1)$$

with energies $E_l = (l+1/2)\omega_0$ where $\omega_0 \sim \Delta_0^2/E_F$ and $\psi_l \sim J_l(k_F r)$ for $r \ll \xi$ (2). Here $J_l$ is the $l^{\text{th}}$ Bessel function of the first kind. The DOS is given by $\rho(\boldsymbol{r}, \omega) = 2 \sum_l |u_l|^2 \delta(\omega - E_l)$ where 2 comes from the two spins. Clearly, the phase information of the CdGM states is lost in the expression of $\rho(\boldsymbol{r}, \omega)$, which gives rise to a continuous circular DOS ring in clean superconductors (Fig. 1, B to D). But in the presence of disorders, the rotational symmetry is broken. In the present sample, there are sparse impurities with strong scattering potentials. They can induce an in-gap impurity bound state and behave as bright dots (20), as shown in Fig. 3A. Besides, there are dense impurities with weak scattering potentials. They cannot induce an apparent in-gap state but act as scattering centers to quasiparticles, forming QPI patterns (fig. S2I) at high energies (20). These dense impurities can break the rotational symmetry, and thus different CdGM states are allowed to interfere with each other, which provides a natural way to detect their phases. Such an idea is quite similar to the widely used technique of QPI in metals (22,23) or superconductors (24). However, the QPI of CdGM states has two additional peculiar properties. First, since the CdGM states are spatially well separated from each other except for the oppositely propagating states (with angular momenta $\pm l$) sharing the same spatial profile (Fig. 1A), the disorder induced interference mainly happens between the two states with angular momenta $\pm l$, and this is a novel manifestation of selection rule. Second, since the $\pm l$-states have an energy difference of $2l\omega_0$, inelastic scatterings are needed to induce the interference. Therefore, for simplicity, we can only consider the interference between $\pm l$-states and call it an off-shell two-level interference provided that the CdGM states are stable against weak disorders. In the first order approximation, the disorder-corrected CdGM states (unnormalized) are given by $\tilde{u}_l = u_l + \alpha_l u_{-l} = \psi_l e^{-il\varphi} + \alpha_l \psi_{-l} e^{il\varphi}$, with the energies $\tilde{E}_l = E_l + |\alpha_l|^2/(E_l - E_{-l})$, where the coupling $\alpha_l = \int \psi_{-l}^* \hat{V}_{\text{imp}} \psi_l d^2 r$ is a complex number reflecting the inelastic scattering between the $\pm l$-states. The resulting local DOS is given by

$$\rho(r, \omega) = 2 \sum_l [1 + |\alpha_l|^2 + 2|\alpha_l| \cos(2l\varphi + \varphi_{0l})] |\psi_l|^2 \delta(\omega - \tilde{E}_l) \quad (2)$$

where $\varphi_{0l}$ is determined by $\alpha_l$. In practice, we can use it to simulate the STM data and $\alpha_l$ is the only fitting parameter, which determines the amplitude and phase of the oscillation. Clearly, there are $2l$ oscillations in the circular direction, and the angular oscillation period is $\Theta_l = 2\pi/2l$. On the other hand, from the radial wave function $\psi_l(r) \sim J_l(k_F r)$, we have an estimation of the radius $R_l \sim l/k_F$, and thus the necklace oscillation wave length $\lambda = R_l \Theta_l \sim \pi/k_F$. This provides an essential understanding of the necklace VBS.

**Fitting results of the experiment data**

The two-level approximation, Eq. (2), allows us to fit the STM data. The radial QP wave functions $\psi_l$ can be calculated by solving the BdG equations (supplementary materials). Figure 3, D to F show the two-level approximation results, which capture the key features of the experimental results. In the above two-level ($\pm l$) analysis, we have ignored the coupling of $|l\rangle$ to adjacent levels $|l \pm 1\rangle$, which will inevitably generate additional harmonics in the wave functions and thus complicate the DOS. In order to justify or go beyond the two-level approximation, we have also performed exact diagonalization studies on disordered lattices. The numerical results are shown in Fig. 3, G to I. Not only the radius of the calculated DOS ring is consistent with the experimental data, but also the DOS oscillations along the angular direction in the calculation results agree well with the experimental results; even some imperfect oscillation features seen in the experimental data can be simulated. Furthermore, the radiative feature of the spots is also well reproduced. At low energies, the radius is too small and the number of spots in the DOS ring is too few, making the oscillation difficult to discern. Nevertheless, it still shows some signs of oscillatory behavior with a similar periodicity. At higher energies near the superconducting gap (Fig. 3C and fig. S2), the DOS shows a diffusive feature. This can be attributed to two reasons. On one hand, more Bogoliubov QPs are excited when the energy is high and the background of normal state DOS becomes prominent. On the other hand, the primary and secondary rings smear together at high energies according to the calculation results.

To further confirm the above theoretical explanation, we make more quantitative comparisons between the experimental and theoretical results. Through analyzing the local DOS distribution of the necklace-like VBS, for example, Fig. 3, A to C and Fig. 4A, we can get the oscillation numbers, namely, $n$, for each VBS. Along the central line of the primary VBS ring (Fig. 4A), the local DOS shows about $n = 17$ complete oscillations (Fig. 4B), and this value is close to the oscillation number $2l = 18$ from the calculations (Fig. 3E). Since the radius $r$ of the VBS ring and $2l$ can characterize the VBS pattern along the radial and angular direction, respectively, we plot these values at different energies and show them in Fig. 4, C and D. The calculation results agree well with the experimental data. This agreement is also observed in several other vortex cores (figs. S2 and S3). Therefore, inversely, one can determine the angular momentum $l$ by simply counting the number of peaks ($2l$). In this way, to the best of our knowledge, our work actually measures the phases (characterized by angular momenta $l$) of the CdGM states experimentally for the first time. The oscillation period is calculated by $2\pi r/2l$, and the obtained periods are 3.9±0.6, 3.8±0.6, and 4.1±1.3 nm at $E = 0.42\Delta_0$, $0.6\Delta_0$, and $0.68\Delta_0$, respectively. These obtained periods are close to the value of $\pi/k_F \approx$ 3.9 nm.

**Phase diagram of the VBS with variable disorders**

In the above, we have learned that the necklace VBS can be explained by disorder-induced interference among different CdGM states. Next, we investigate the effect of disorder strengths and densities. For a randomly distributed disorder $[-V_\text{imp}, V_\text{imp}]$ with the density of $n_\text{imp}$, the coupling strength $\alpha_l$ can be statistically estimated as $\alpha_l \propto n_\text{imp}\sqrt{V_\text{imp}^2/n_\text{imp}} \propto \sqrt{\Gamma_\text{imp}}$, where $\Gamma_\text{imp} \propto n_\text{imp}V_\text{imp}^2$ is the disorder scattering rate. According to Eq. (2), we need $\Gamma_\text{imp}$ to be large enough (not in the clean limit) to observe the oscillation behavior. On the other hand, $\alpha_l$ also shifts the energy level $E_l$. When the shift is much larger than the energy spacing between adjacent CdGM states, the original CdGM states are expected to be destroyed, and hence the above analysis will break down. This provides an upper bound of $\Gamma_\text{imp}$ (not in the dirty limit) to observe the necklace rings. The above analyses are checked by numerical calculations systematically. The main results are summarized in Fig. 5. The phase diagram versus $n_\text{imp}$ and $V_\text{imp}$ is roughly divided into three regimes, labeled as I, II, and III, respectively (Fig. 5A). Typical local DOS of the three regimes are shown in the insets. With increasing the scattering rate $\Gamma_\text{imp}$, the continuous ring (in regime I) is firstly broken into the necklace shape (in regime II), and finally destroyed (in regime III). From the angular distribution of local DOS along the primary rings (Fig. 5B), we can see a transition from regime I to III with increasing $\Gamma_{imp}$. More detailed results about the phase diagram are shown in figs. S6 and S7. The results in regime III are well supported by the STM measurements of Ni-doped K12442 samples, where the disorder scattering becomes strong, and the VBS do not show a well-formed necklace pattern (*25*). According to our calculation (figs. S8 and S9), the situation for clearly observing the discrete CdGM states is different from that of seeing the necklace vortex pattern. In the former case, one needs a clean system with a very small disorder scattering rate $\Gamma_\text{imp}$; while in the latter case, an appropriate disorder scattering rate is required.

**Discussion and conclusion**

Our experiments reveal that some vortex patterns exhibit as a little squarish shape at high energies (Fig. 2 and Fig. 3C), and the sides of the squarish vortices are approximately along the crystalline axes of the reconstructed surface by K or Ca atoms (same as the Fe-Fe direction). Some DOS modulation spots in the side regions of the squarish vortex are roughly oriented along two orthogonal directions. However, in the corner regions of the square-shaped vortex, the DOS modulations orient clearly in the diagonal directions showing a radiative feature, which deviates from the above-mentioned two orthogonal directions. Therefore, the spots of DOS modulations show a radiative feature with orientations always perpendicular to the contour of the DOS ring.

One may argue that the necklace-like pattern of the VBS may be caused by the anisotropic superconducting gap or the anisotropic Fermi surface, which can break the rotational symmetry of the vortex core (*26,27*). In these cases, vortex cores may exhibit twofold, fourfold, or sixfold symmetric shapes following the rotational symmetry of $\Delta_0$ or $v_F$ (*28-37*). The vortex-core patterns expand in size with increasing energy, but the shape and symmetry do not change clearly (*31,32*). There are no multiple oscillations of DOS with the periodicity of $\pi/k_F$ along the angular direction. In order to check whether the anisotropy in $v_F$ can induce the oscillation of VBS along the angular direction, we calculate the VBS by solving the BdG equations in a clean system with a fourfold symmetric Fermi surface (fig. S10), from which one can see the squarish vortex core pattern especially at high energies, but we do not see any signatures of the DOS modulations along the contour of the VBS ring. In addition, from our experiments, we can easily see that the vortex patterns at low energies exhibit a roughly round shape with DOS modulation spots along the radiative direction (Fig. 1, F and G, Fig. 3, A and B). This can exclude the possibility that the DOS modulation along the angular direction is induced by the anisotropic Fermi velocity or the gap anisotropy.

As we mentioned above, this exotic pattern of VBS has been observed in the present iron-based superconducting system with a small Fermi energy $E_F$ or a large Fermi wavelength $\lambda_F$, while in the Ni-doped K12442 with a larger residual resistivity (higher disorder density or stronger scattering potential), this phenomenon was not observed (*25*). In short, to observe the circular oscillatory VBS rings, we need a small $k_F$ (large $\lambda_F$) and an appropriate disorder scattering rate $\Gamma_{imp}$, which are not easy to satisfy in most superconductors. However, these conditions are exactly satisfied in some iron-based superconductors, like our present system K12442. Our observation of the novel necklace-like vortex pattern may inspire researchers to see whether such a phenomenon also exists in topological non-trivial vortex with Majorana zero mode and other interger-quantized VBS, and the recent data seem to show some feature of this (*17*).

In summary, we discovered an unprecedented necklace-like VBS in iron-based superconductor K12442 using STM. In contrast to the continuous DOS ring-like pattern predicted by the BdG equations in the clean limit, here we find a multiple oscillatory feature of VBS along the angular direction, which has never been observed before and predicted by any previous theoretical studies. The DOS oscillation period is found to be close to $\pi/k_F$. In order to interpret the observation, we propose a model concerning a selective off-shell interference between CdGM states with opposite angular momenta, enabled by the wave-function matching and rotational symmetry breaking due to disorders. The numerical calculations based on this model reproduce nicely the necklace-like VBS, which can be most favorably observed in a system with a small Fermi wave vector $k_F$ and appropriate disorder scattering rate $\Gamma_{imp}$. This is a significant discovery which provides a new way to detect the phase of the CdGM states experimentally. Our results shed new light on the understanding of the nature of QPs and vortices in superconductors with a small Fermi energy, and will stimulate further interest in experimental and theoretical studies.

**Acknowledgments:** We acknowledge helpful discussions with Hong Ding, Christophe Berthod and Egor Babaev. **Funding:** This work was supported by the National Key R&D Program of China (grants 2022YFA1403200, 2018YFA0704200, 2023YFA1406100, 2022YFA1403400, and 2021YFA1400400), National Natural Science Foundation of China (grants 12061131001, 11927809, 11888101, 12374147 and 12274205), the Strategic Priority Research Program (B) of the Chinese Academy of Sciences (grants XDB33000000, GJTD-2020-01), and the Youth Innovation Promotion Association of the Chinese Academy of Sciences (grant Y202001). **Author contributions:** Single crystal synthesis: WH, SL, HL; STM/STS measurements and analysis: KC, ZH, WD, HY, HHW; Theoretical calculation: DW, QW, ZH, TX; Writing: ZH, HY, HHW, DW, TX; Coordinated the whole work: HHW. **Competing interests:** The authors declare that they have no competing interests. **Data and materials availability:** All data needed to evaluate the conclusions in the paper are present in the paper and the Supplementary Materials. Additional data related to this paper may be requested from the authors.

**Supplementary Materials**
Materials and Methods
Supplementary Text
Figs. S1 to S10

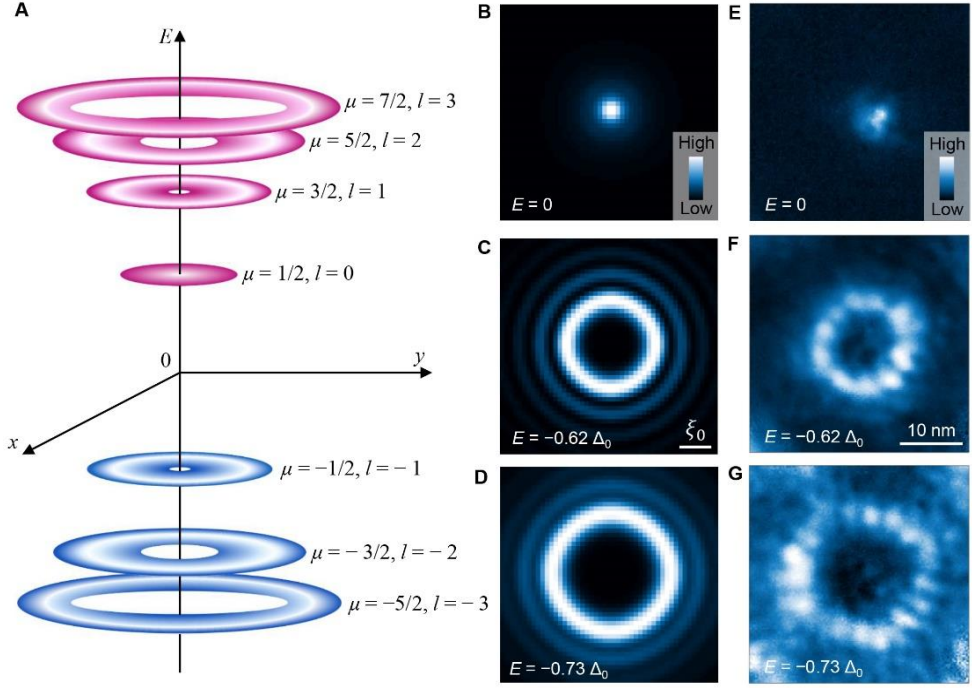

**Fig. 1. CdGM states and necklace-like vortex bound states.** (**A**) Schematic diagram of discrete VBS energy levels and the DOS rings based on the calculations of the BdG equations. (**B**-**D**) Evolution of local DOS of the calculated CdGM states with energy. $\Delta_0$ is the superconducting gap far away from the vortex core, and the parameter $k_\mathrm{F}\xi = 5$. (**E**-**G**) The d$I$/d$V$ mappings of a necklace-like vortex core measured at different energies inside the superconducting gap.

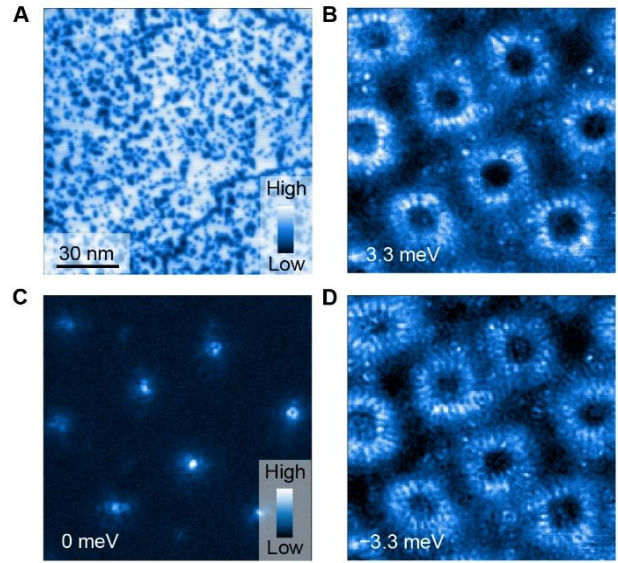

**Fig. 2. d*I*/d*V* mappings of multiple vortices with necklace-like VBS in a large area.**
(**A**) STM topography of KCa$_2$Fe$_4$As$_4$F$_2$. (**B-D**) d*I*/d*V* mappings measured at different energies ($\mu_0 H$ = 2 T). The measuring region is the same as (A).

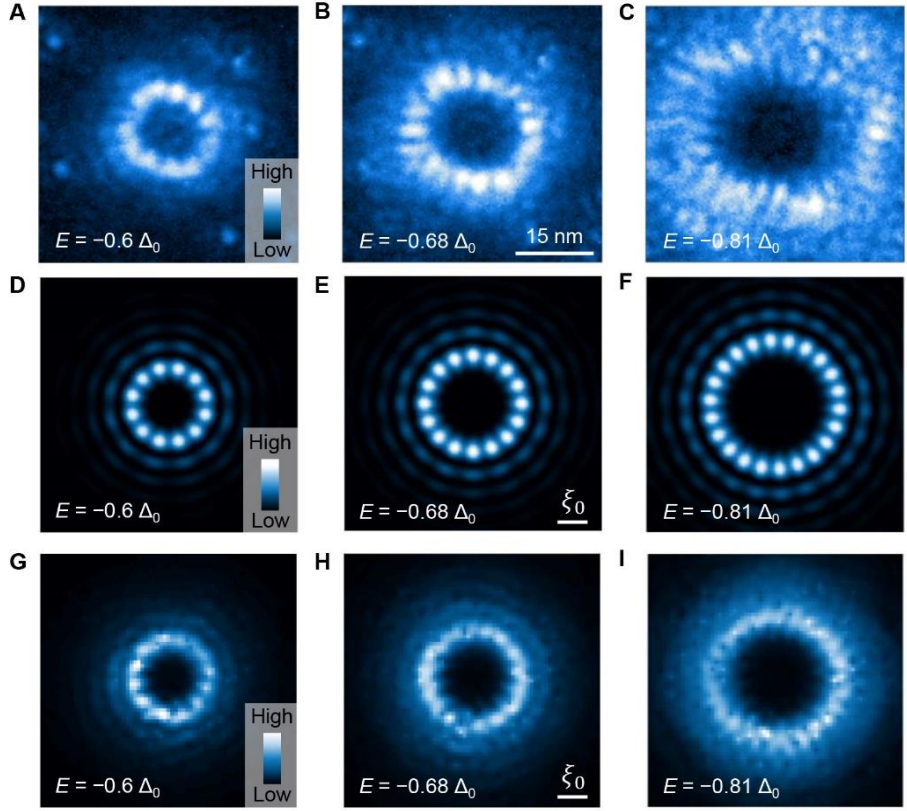

**Fig. 3. Isolated vortex core and calculation results by using the disorder-corrected CdGM states.** (**A**-**C**) d$I$/d$V$ mappings of a necklace-like vortex core measured at different energies ($\mu_0 H$ = 0.2 T). (**D**-**F**) Two-level approximation with the parameter $k_F \xi$ = 5.7. Here, $k_F$ is set to the similar value as the experimental data, and the theoretical patterns are plotted with the same scale bar as the experimental data. (**G**-**I**) Numerical calculation results by exact diagonalization of the disorder-corrected CdGM states ($n_{imp}$ = 10% of all Fe sites, and $V_{imp}$ = 0.36$\Delta_0$).

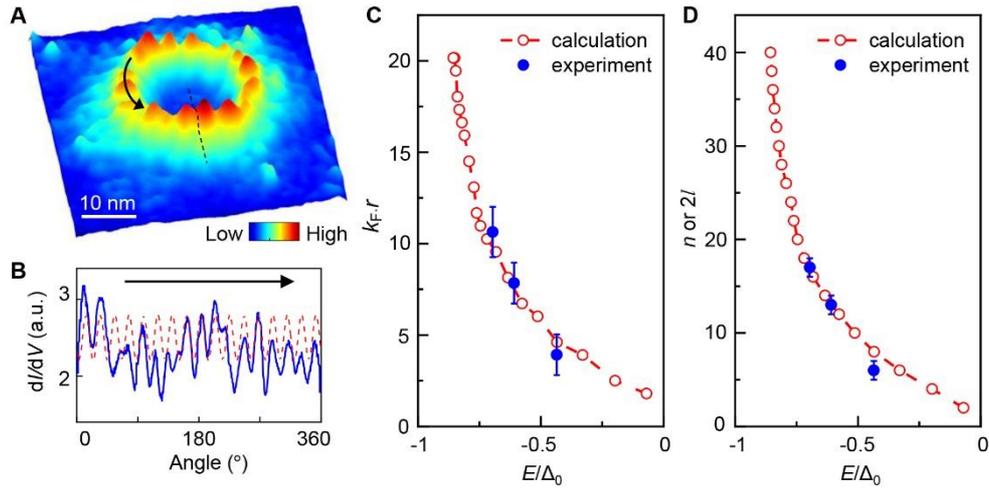

**Fig. 4. Comparison of experimental and calculation results.** (**A**) A representative d$I$/d$V$ mapping of necklace-like VBS pattern plotted in a three-dimensional manner after smoothed. (**B**) The distribution of d$I$/d$V$ intensity after smoothing along the DOS ring marked by the black arrow in (A). The initial angle is from the dashed line in (A). The red dashed curve is a sinusoidal function with 17 complete oscillations in a circle. (**C**) Comparison of the radii of VBS rings acquired from experiment (solid circles) and calculation (empty circles). (**D**) Comparison of the number of peaks in the DOS ring, i.e., $n$ for experiment (solid circles) and $2l$ for calculation (empty circles). The experimental results consist well with the calculation result with $k_F\xi = 5.7$. The error bars in (C) and (D) are determined by standard errors of the mean values of the radii and the uncertainty in counting the oscillation numbers, respectively.

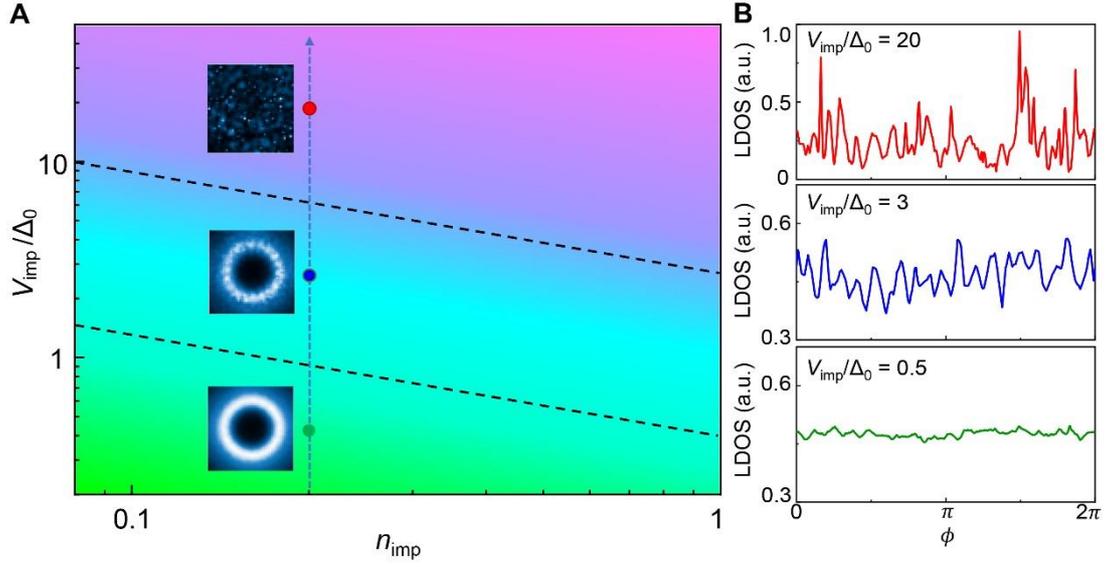

**Fig. 5. Phase diagram of the VBS.** (**A**) The phase diagram on the plane of ($n_{imp}$, $V_{imp}$) contains three regimes: I (ring), II (necklace), and III (featureless). Typical local DOS distributions of the three regimes are shown in the insets. The dark dashed lines show the approximate boundaries between neighbored regimes. (**B**) Angular distributions of the DOS along the same ring for different $V_{imp}$. As $V_{imp}$ increases, the curves behave differently from regime I to III.

# Supplementary Materials for

## Necklace-like pattern of vortex bound states

Zhiyong Hou[1†], Kailun Chen[1†], Wenshan Hong[2,3†], Da Wang[1†], Wen Duan[1], Huan Yang[1*], Shiliang Li[2,4,5], Huiqian Luo[2,5], Qiang-Hua Wang[1*], Tao Xiang[2,4,6], Hai-Hu Wen[1*]

[*]Corresponding author. Email: huanyang@nju.edu.cn (H. Y.); qhwang@nju.edu.cn (Q.-H.W.); hhwen@nju.edu.cn (H.-H.W.)

**The file includes:**
Materials and Methods
Supplementary Text
Figs. S1 to S10

## 1. Materials and Methods

The KCa$_2$Fe$_4$As$_4$F$_2$ (K12442) single crystals were grown by the self-flux method (*38*). A scanning tunneling microscope (USM-1300, Unisoku Co., Ltd.) was used to carry out the STM/STS measurements. The samples of K12442 were cleaved at about 77 K in an ultrahigh vacuum with a base pressure of about 1×10$^{-10}$ Torr, and then transferred to the STM head. Electrochemically etched tungsten tips were cleaned by electron-beam heating and then used for STM/STS measurements. A typical lock-in technique was used in tunneling spectrum measurements with an ac modulation of 0.1 mV and a frequency of 931.773 Hz. Setpoint conditions for tunneling spectrum measurements are $V_{set}$ = 10 mV and $I_{set}$ = 200 pA. All the experimental data were measured at about 0.4 K.

## 2. Calculations on a lattice model with disorders

We have performed numerical calculations on a square lattice with the Hamiltonian given by

$$H = -t \sum_{\langle ij\rangle\sigma} \left(c_{i\sigma}^\dagger c_{j\sigma} + h.c.\right) + \sum_{i\sigma}(V_i - \mu_c)c_{i\sigma}^\dagger c_{i\sigma} + \sum_i \left(\Delta_i c_{i\uparrow}^\dagger c_{i\downarrow}^\dagger + h.c.\right), \quad (S1)$$

where $c_{i\sigma}^\dagger$ creates an electron on site $i$ with spin $\sigma$, $t$ is the nearest neighbor hopping taken as the energy unit, $\mu_c$ is the chemical potential. The pairing function is approximated as $\Delta_i = \Delta_0 \tanh(r_i/\xi) e^{i\phi_i}$ where $(r_i, \phi_i)$ are polar coordinates of the site $i$ relative to the vortex center. We have added scalar disorder potentials $V_i$ following a uniform distribution [$-V_{imp}$, $V_{imp}$] and the density $n_{imp}$ defined as the ratio between the impurity sites and Fe atoms in a layer. By exact diagonalization, the low energy bound states, $E_n$ and $(u_n, v_n)$, are obtained on the $L \times L$ lattice, from which the local DOS is given by

$$\rho(i,\omega) = \frac{2}{\pi} \sum_n \frac{|u_n(i)|^2 \eta}{(\omega - E_n)^2 + \eta^2}, \quad (S2)$$

where $\eta$ is the smearing factor. In calculations, unless specified, we set $\mu_c = -3.5$, $\Delta_0 = 0.1$, $\xi = 10$, $L = 200$, $\eta = 0.004$. By these parameters, we have $k_F\xi \approx 7$, which is similar to the Γ-pocket in KCa$_2$Fe$_4$As$_4$F$_2$.

For $n_{imp}$ = 0.2 and $V_{imp}$ = 3$\Delta_0$, we plot the wave functions $|u_l|^2$ with $l$ = 0, …, 15 in fig. S6. The disorder-induced oscillation can be seen for all the states except $l$ = 0. Of course, for the several low energy states, since the numbers of necklace oscillations are small, they are easy to be broken by disorder and difficult to identify accurately. On the other hand, for high energy states, the oscillation is clearly seen and one can always perform Fourier transformation to extract the leading oscillation period. Next, we scan the ($n_{imp}$, $V_{imp}$) plane to study the condition to observe the necklace-like behavior. The results of $|u_6|^2$ for a series of ($n_{imp}$, $V_{imp}$) are shown in fig. S7. Clearly, the complete ring is firstly broken into the necklace shape and finally destroyed by increasing either $n_{imp}$ or $V_{imp}$. These results can be roughly understood as explained in the main text: a single parameter $\Gamma_{imp} \propto n_{imp} V_{imp}^2$ can be used to describe the vortex bound states from clean limit (ring) to dirty limit (none), and the necklace-like pattern occurs in between.

To see the effect of the Fermi surface, we tune $\mu_c = -1.0$ to simulate a clean system with a squarish Fermi surface. The results are shown in fig. S10. The quantized VBS states are obtained. The DOS mappings for several energies are presented, showing a four-fold feature but the oscillation behavior is not seen, indicating the Fermi surface shape is not the reason for the necklace feature.

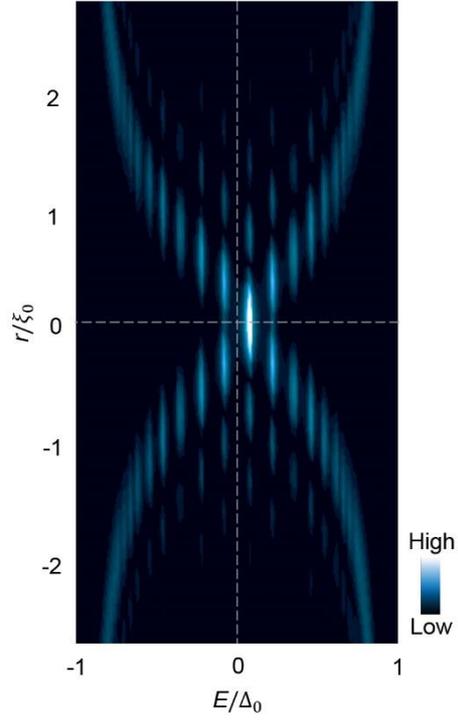

**Fig. S1. Spatial and energy evolution of the local density of states calculated for a usual vortex.** Spatial evolution of the local density of states calculated at different energies. The calculation is carried out with $T = 0.01T_c$ and $k_F\xi = 5$.

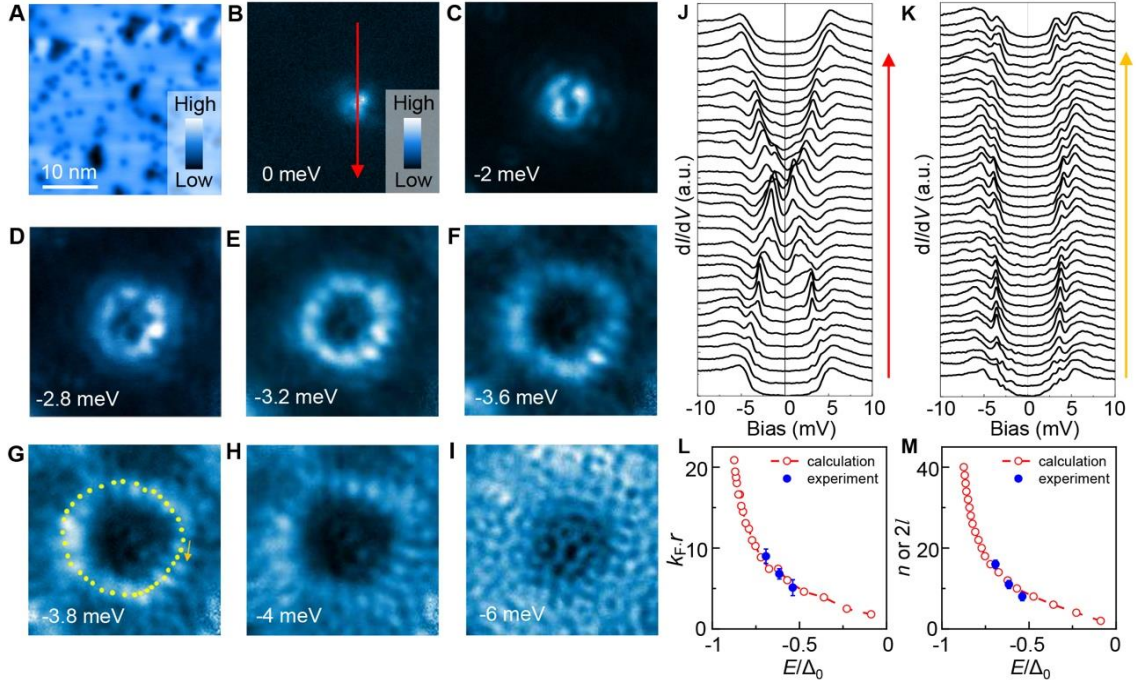

**Fig. S2. Evolution of necklace-like VBS with energy.** (**A**) Topography of an area where a single necklace-like vortex is located. (**B**) The d$I$/d$V$ mapping of the same field of view measured at 0 meV ($\mu_0 H = 2$ T). (**C-H**) The d$I$/d$V$ mappings measured at different energies inside the superconducting gap and (**I**) outside the gap. When the energy exceeds the superconducting gap, all QPs are excited and the background DOS around the vortex becomes dominant, forming a standing wave pattern induced by QPI of the CdGM states. (**J**) Tunneling spectra measured along the red arrow in (B). (**K**) Tunneling spectra measured along the DOS ring as marked by yellow dots in (G). (**L**) Comparison of the radii acquired from experiment (solid circles) and calculation (empty circles). (**M**) Comparison of the number of peaks in the DOS ring, i.e., $n$ for experiment (solid circles) and $2l$ for calculation (empty circles). The calculation parameter is $k_F \xi = 5$. The error bars in (L) and (M) are determined by standard errors of the mean values of the radii and the uncertainty in counting the oscillation numbers, respectively.

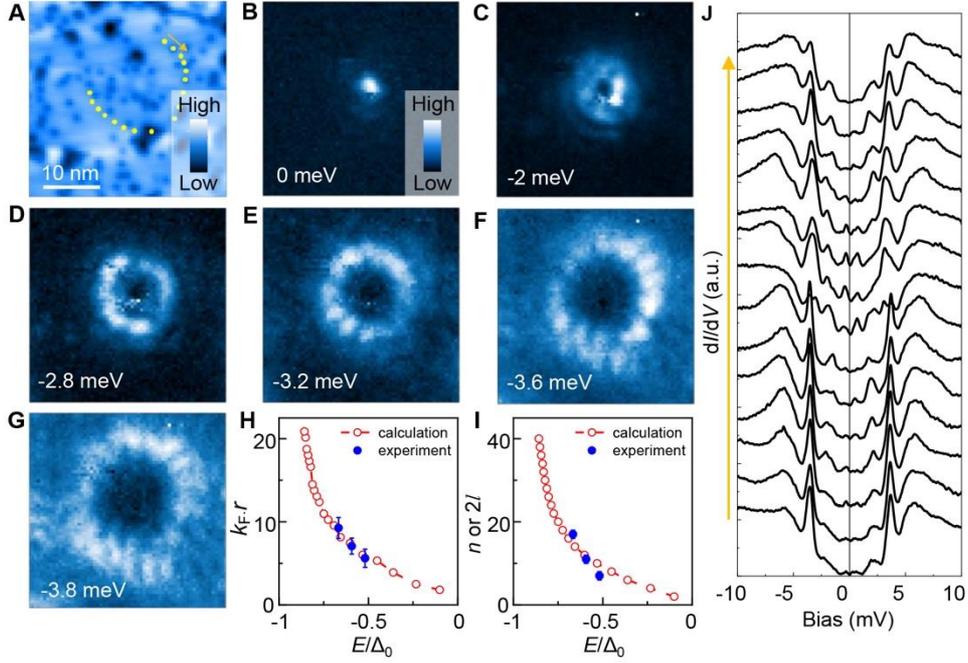

**Fig. S3. Discrete energy levels of a necklace-like vortex.** (**A**) Topography of a necklace-like vortex location. (**B-G**) The d$I$/d$V$ mappings measured at different energies ($\mu_0 H$ = 2 T). (**H**) Comparison of the radii acquired from experiment (solid circles) and calculation (empty circles). (**I**) Comparison of the number of peaks in the DOS ring, i.e., $n$ for experiment (solid circles) and $2l$ for calculation (empty circles). The calculation parameter is $k_F \xi$ = 5. The error bars in (H) and (I) are determined by standard errors of the mean values of the radii and the uncertainty in counting the oscillation numbers, respectively. (**J**) Tunneling spectra measure along the DOS ring as marked in (A). Discrete in-gap bound states exhibited as differential conductance peaks in the spectra show up clearly and the highest peak amplitude appears at ±3.6 meV.

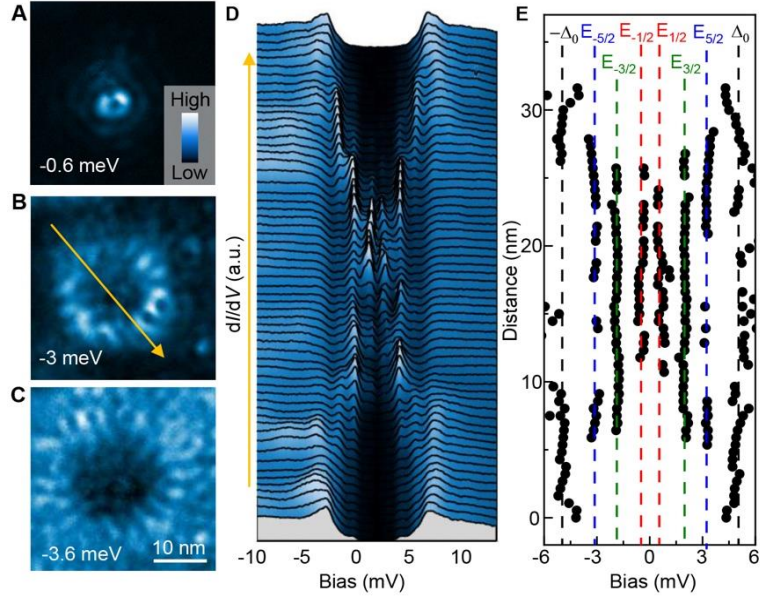

**Fig. S4. Spatial variations of discrete VBS in necklace-like vortex.** (**A-C**) The d$I$/d$V$ mappings measured at different energies for a necklace-like vortex ($\mu_0 H$ = 2 T). (**D**) Tunneling spectra measured across the vortex, as marked by the arrow in (B). (**E**) Spatial variation of the bound-state energy derived from (D). The first three energy levels ($\mu = \pm 1/2, \pm 3/2, \pm 5/2$) of the VBS can be easily distinguished.

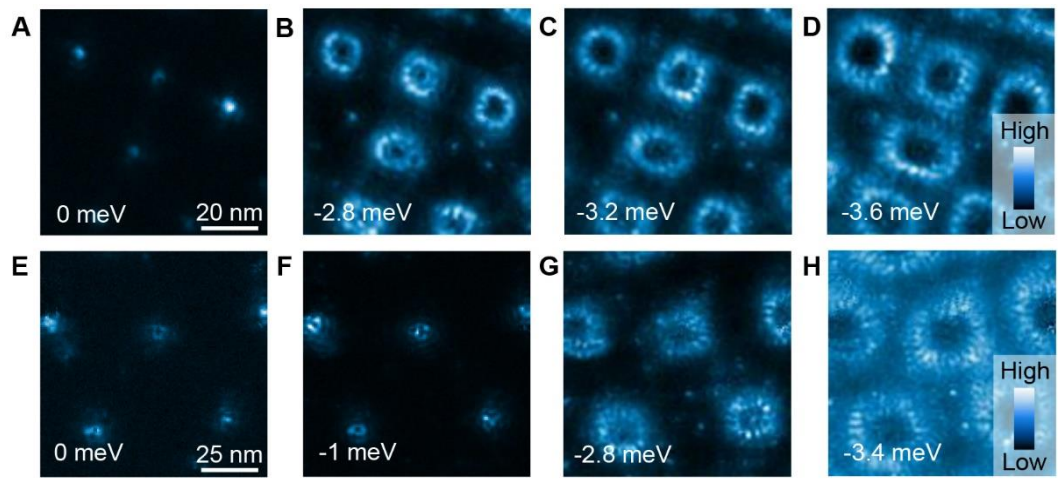

**Fig. S5. Multiple necklace-like VBS patterns under a magnetic field of 2 T.** (**A-D**) and (**E-H**) show the d$I$/d$V$ mappings of two different areas with multiple necklace-like vortices.

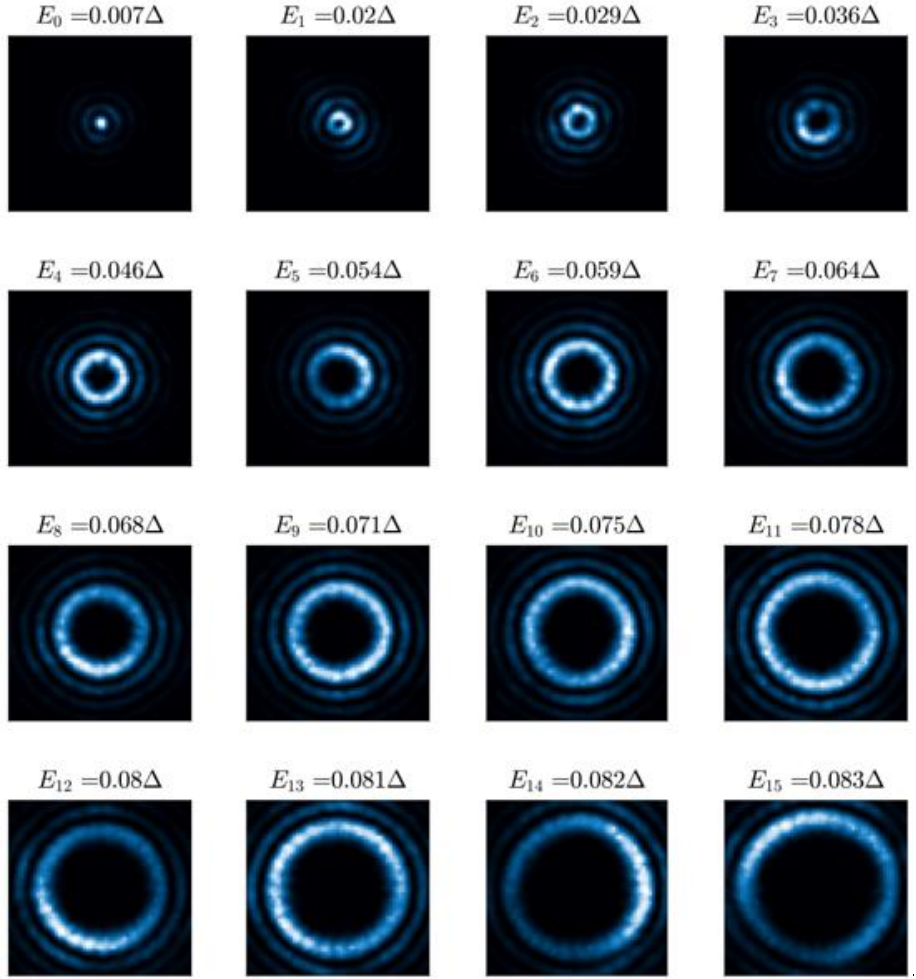

**Fig. S6. VBS wave functions $|u_l|^2$ ($l$ = 0, 1, …, 15) for $n_{imp}$ = 20% of all Fe sites and $V_{imp}$ = 3$\Delta_0$.**

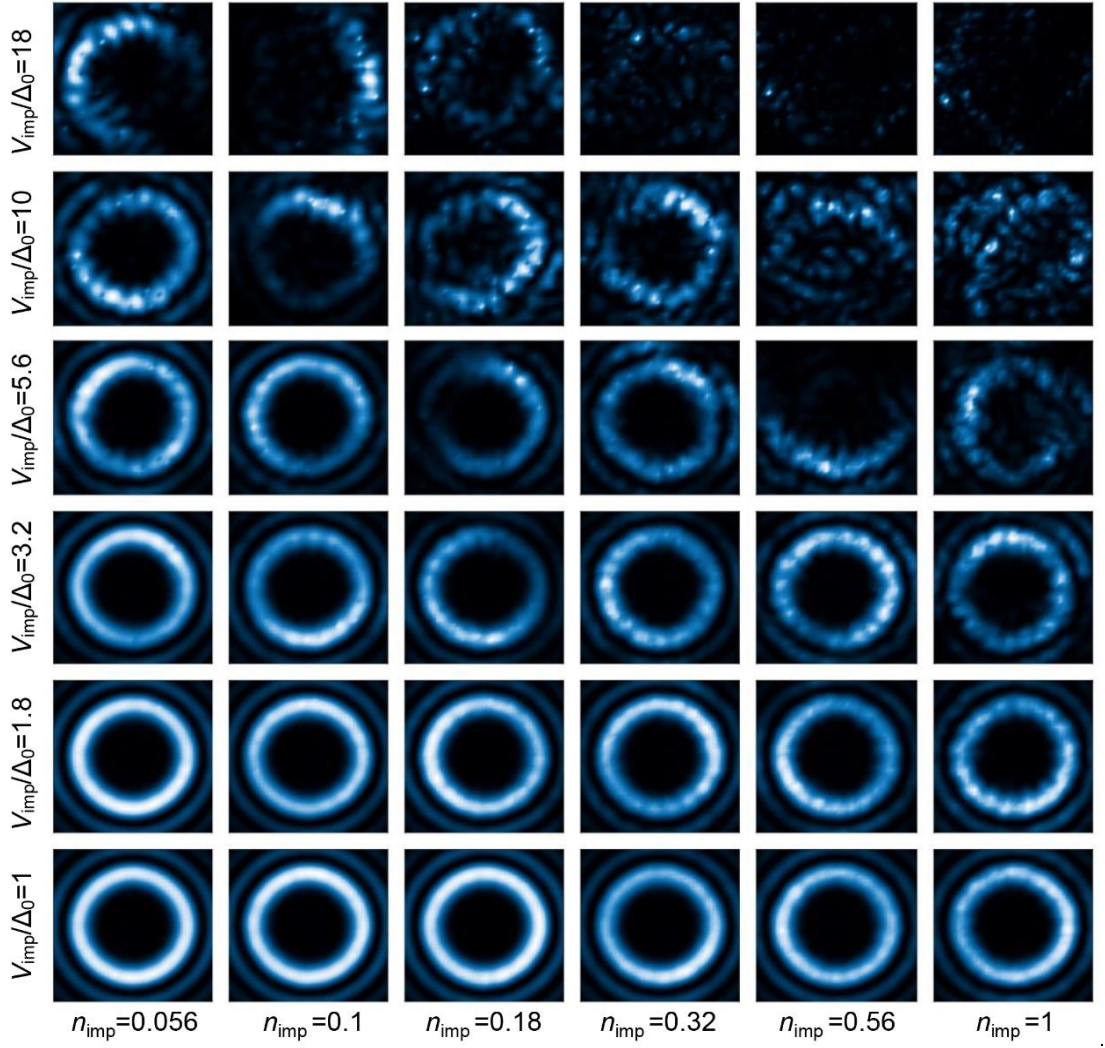

**Fig. S7. VBS wave functions $|u_6|^2$ calculated for different values of ($n_{imp}$, $V_{imp}$).**

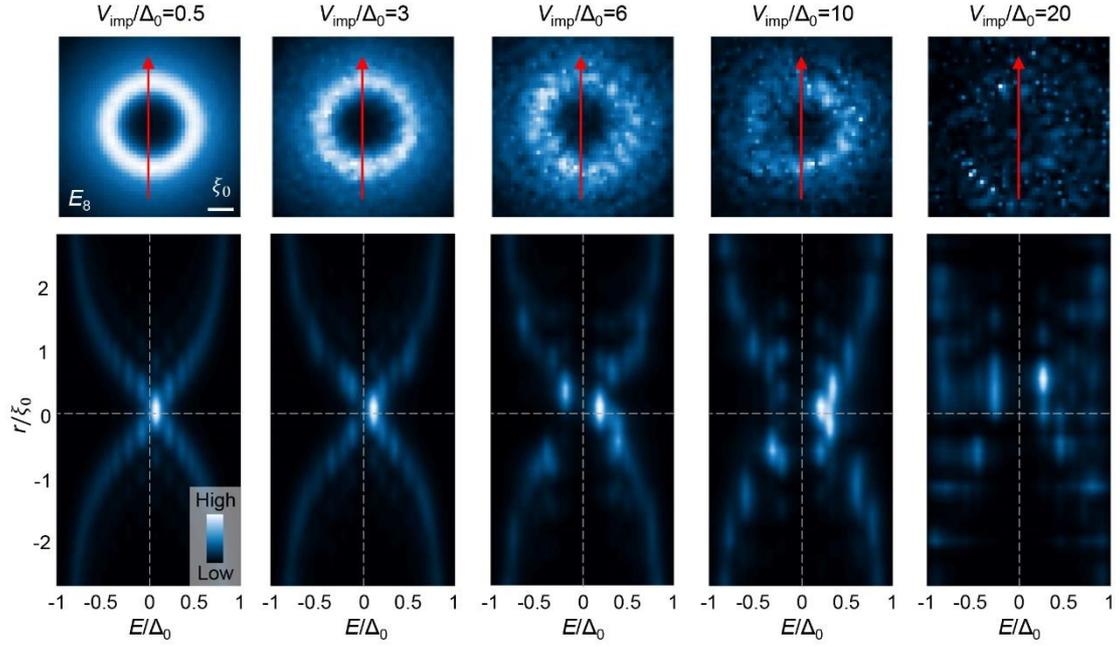

**Fig. S8. Calculated local DOS across vortex core. Upper row:** Calculated DOS at the energy of $E_8$ for different disorder potential $V_{imp}$. The impurity density is set to be $n_{imp} = 20\%$ of all Fe sites, and $k_F\xi = 5$. **Lower row**: Two dimentional plot of the spatial and energy dependent of DOS along the red arrows in the upper row. The dense strong disorders also affect the CdGM bound state, including the spatial- and energy-dependence of the CdGM bound state.

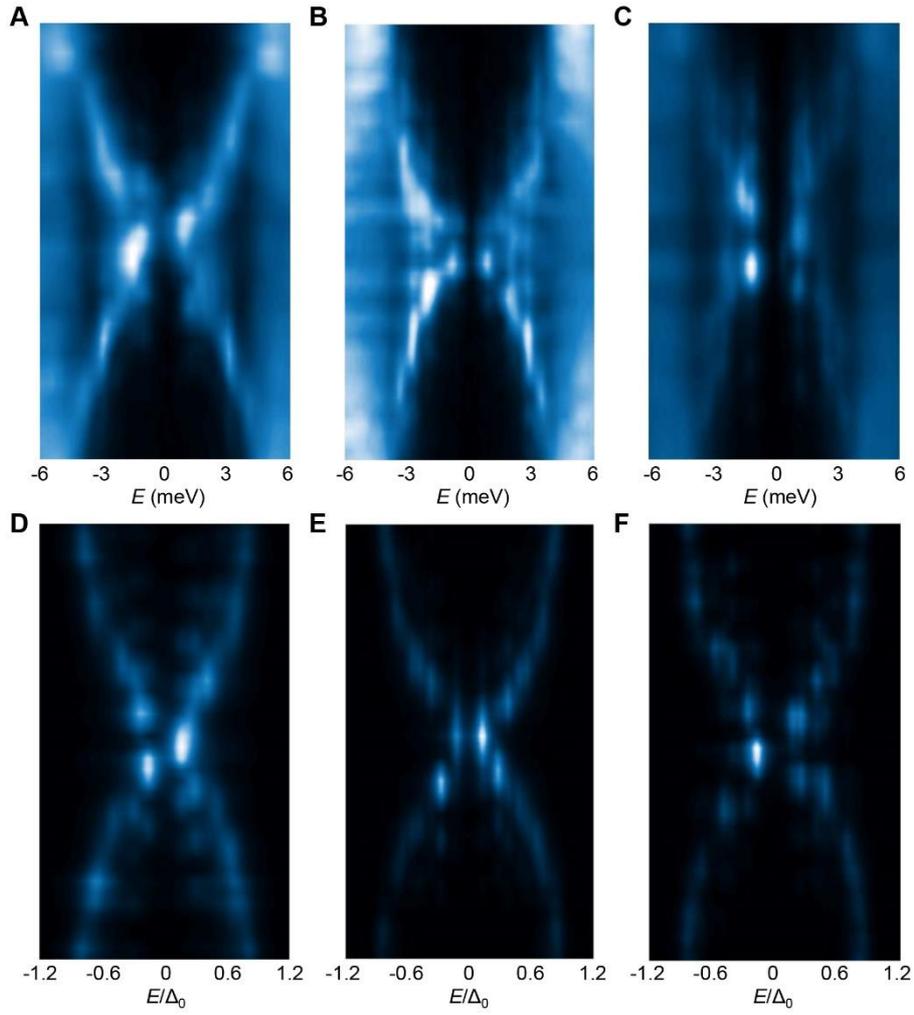

**Fig. S9. Measured and calculated tunneling spectra across vortex cores.** (**A-C**) Different tunneling spectra measured across vortex cores for the same K12442 sample. (**D-F**) Calculation results of spatial evolution of local DOS in the presence of disorders with suitable density and scattering potentials.

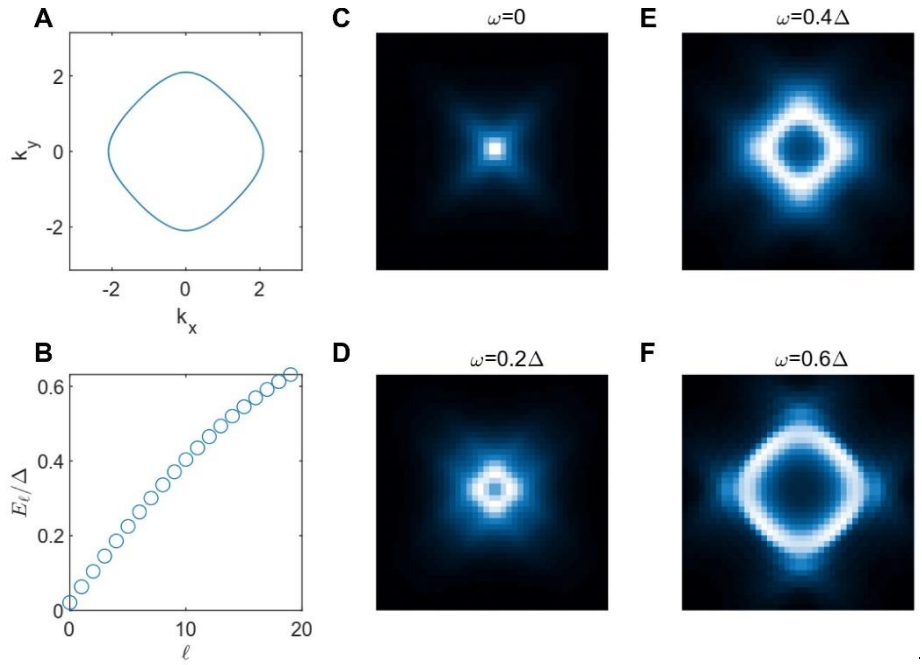

**Fig. S10. Calculated VBS results for a squarish Fermi surface.** (**A**) Schematic diagram of the squarish Fermi surface. (**B**) Calculated VBS energy levels. (**C-F**) Calculated local DOS distributions at different energies.